\documentclass[useAMS,usenatbib]{mn2e}
\usepackage{graphicx}
\usepackage{amsmath}
\usepackage{amssymb}

\title[M82 ULX millisecond pulsar]{An ultraluminous nascent millisecond pulsar}
\author[W. Klu\'zniak, J.-P. Lasota]{W{\l}odek Klu\'zniak$^{1}$\thanks{E-mail:
wlodek@camk.edu.pl} and  Jean-Pierre Lasota$^{2,1,3}$
\thanks{E-mail:lasota@iap.fr}\\
$^{1}$ Copernicus Astronomical Center,
 ul. Bartycka 18, PL 00-716 Warszawa, Poland\\
$^{2}$ CNRS, UMR 7095, Institut d'Astrophysique de Paris, 98bis Bd Arago, 75014 Paris, France\\
$^{3}$ Sorbonne Universit\'es, UPMC Univ Paris 06, UMR 7095, 98bis Bd Arago, 75014 Paris, France \\}
\begin{document}

\date{Accepted 2014 December 11. Received 2014 November 24; in original form 2014 November 4}

\pubyear{448, L43toL47, 2015}

\maketitle

\label{firstpage}

\begin{abstract}
If the ultraluminous source (ULX) M82 X-2 sustains its measured
spin-up value of $\dot \nu= 10^{-10}\,{\rm s^{-2}}$, it will become a
millisecond pulsar in less than $10^5\,$ yr. The
observed (isotropic) luminosity of $10^{40}\,$ erg/s also supports the
notion that the neutron star will spin up to a millisecond period upon
accreting about $0.1\,{\rm M_\odot}$---the reported hard X-ray luminosity of this
ULX, together with the spin-up value, implies torques consistent with
the accretion disc  extending down to the vicinity of the stellar
surface, as expected for low
values of the stellar dipole magnetic field ($B\lesssim 10^9\,$G). 
  This suggests a new channel of
millisecond pulsar formation---in high-mass X-ray binaries (HMXBs)---and
may have implications for studies of gravitational waves, and possibly
for the formation of low-mass black holes through accretion-induced
collapse.
\end{abstract}

\begin{keywords}
accretion: accretion discs - gravitational waves - magnetic fields - 
stars: black holes - stars: neutron - pulsars: NuSTAR J095551+6940.8
\end{keywords}
\section{INTRODUCTION}

The unexpected discovery of a 1.37 s pulsation in the M82 X-2 ultraluminous
source \citep{Bachettietal14} invites a reevaluation of common assumptions
about the nature of ULXs and the evolutionary paths of pulsars.

ULXs were universally considered to be accreting black holes,
most likely $\sim10 \,{\rm M_\odot}$ ones in a binary with a high-mass companion,
with some sources possibly harbouring
 ``intermediate mass''  $\sim10^{2-3} \,{\rm M_\odot}$ 
black holes \citep{Kingetal01,Roberts07}.
Their high luminosity was thought to imply that if the X-ray source is
of typical stellar mass, the emission should be beamed, perhaps in
an accretion funnel.

{\sl Chandra} discovery of a population of ULXs  with lifetimes
$\lesssim 10^7$ yr in the Cartwheel galaxy suggests that ULXs
should be HMXBs \citep{King04}, so $\sim 10\, \,{\rm M_\odot}$
 black holes seemed to be the preferred
model. However, not all HMXBs are black hole systems:
(non-ULX) binaries composed of a neutron star
and a massive stellar companion are routinely observed,
both in detached systems \citep[e.g., Be binaries;][]{Reig11},
 and semi-detached 
systems \citep[accretion powered X-ray pulsars;][]{vPMc95}. 

\citet{Bachettietal14} report a $P=1.37\,$s pulsar
in M82 X-2, which is spinning up at the rate $\dot P=-2\cdot 10^{-10}$,
and whose emission shows a 2.5 d sinusoidal modulation,
interpreted as the orbital motion of the X-ray source revolving
around a $>5 \,{\rm M_\odot}$ companion. 
Clearly, this ULX is a neutron star HMXB.
Only a few weeks earlier, a nearby source M82 X-1 was reported
to exhibit a 5 Hz frequency, interpreted at the time as a high frequency
QPO in a $\sim 400\,{\rm M_\odot}$ black hole \citep{Pasham14};
one wonders whether the 5~Hz frequency could instead have been a
harmonic of a 0.6~s pulsar, so that both ULXs in M82 might be harbouring
a neutron star. If, instead, M82 X-1 is indeed an intermediate mass black hole,
one would expect the similarity of X-ray properties of the two sources
to imply that the non-pulsed emission from M82 X-2 originates
in the accretion disc, as it must in the (presumed) black hole M82 X-1.
 
In their discussion \citet{Bachettietal14} assume the pulsar
NuSTAR J095551+6940.8 to be a run-of-the-mill
accretion-powered neutron star pulsar endowed with a $10^{12}\,$G
magnetic field (dipole moment $\mu=10^{30}\,{\rm G\cdot cm^3}$).
In our view, such an assumption is difficult to reconcile with the data.
A dipole moment this strong would disrupt the accretion disc at a large
distance (about 100 stellar radii) from the neutron star. 
However, while the measured spin-up rate is uncommonly high,
the ratio of the measured value of the frequency derivative to
the luminosity ($L_X\approx10^{40}\,$erg/s) is incompatible
with a large lever arm of the accretion torque 
typical of X-ray pulsars (such as Her X-1), where the lever arm
corresponds to the magnetic radius $\sim10^8-10^9\,$cm.
Moreover, for an inner
disc  radius so large, the disc
emission would be in soft X-rays, implying that on this hypothesis
the hard X-ray luminosity observed by Nu-STAR  cannot originate
in the accretion disc -- the observed luminosity would have
to originate in the polar accretion column
close to the neutron star surface, making 
any similarity of the unpulsed X-ray emission to that of black-hole ULXs
purely fortuitous. 

Future observations will show whether the current spin-up rate
is secular. However, at present there is no compelling reason
to assume that the large spin-up torques present in the system today
are going to be displaced by equally large spin-down torques in the
foreseeable future. If the current properties of the system were
extrapolated into the future, one would conclude that the pulsar will
be spun up to millisecond periods already upon accreting 
$\sim 0.1 \,{\rm M_\odot}$,
which is a small fraction of the donor star's mass.

 At the current spin-up rate, $\dot \nu=-\dot P/P^2=10^{-10}\,{\rm s^{-2}}$,
the ultraluminous pulsar would become a millisecond pulsar
in less than 100 000 years:
$\nu =T\dot \nu=300\,$Hz in $T= 10^5\,$y.
Subsequent evolution
of the system depends on how successful the neutron star is in expelling
the mass and angular momentum transferred from the companion -- the system
could become a millisecond accreting pulsar, it could become a radio pulsar
ablating its companion (a ``black widow'' progenitor), or it could end as
a black hole binary, possibly a ULX, upon accretion induced collapse of the
neutron star.

\section{spin-up of the ultraluminous pulsar M82 X-2}
\label{s2}

A remarkable feature of ULXs is, of course, their large luminosity.
It is even more remarkable now that the compact source has been identified
as a pulsar, as this would imply a $\sim 1.4 \,{\rm M_\odot}$ mass and an
apparent (isotropic) luminosity a hundred times the Eddington value,
$L_X\approx 10^2 L_{\rm Edd}$.
However, for this ULX pulsar the most striking feature is
its measured spin-up rate.
 On the one hand, in absolute terms, the spin-up
rate $\dot \nu=10^{-10}\,{\rm s^{-2}}$
 is orders of
magnitude higher than the values measured in the usual accretion powered
 X-ray pulsars,
e.g.,   $3.7\times 10^{-13}\,{\rm s^{-2}}$ in Her X-1,
or $7\times 10^{-12}\,{\rm s^{-2}}$ in Cen X-3
\citep{Bildstenetal97,1985AcA....35..185Z}.
On the other hand, the spin-up to luminosity ratio 
$10^{-50}\,{\rm (erg\cdot s)}^{-1}$
is an order of magnitude lower than the typical ratio observed
in the X-ray pulsars. 

It is this ratio, $\dot \nu/L_X=10^{-50}\,{\rm (erg\cdot s)}^{-1}$,
which makes an interpretation of the data in terms of a strongly
magnetized X-ray pulsar quite challenging.  One would need to find a
model in which the accretion disc, even though truncated\footnote{So that, according to
 the criticized model, in the near future the torque
  on the neutron star could undergo a reversal of sign owing to
  magnetic interactions \citep[][and references
  therein]{Pringle72,Rappaport77,KR07}.} close
  to  the co-rotation radius at the current 
  period (at $r\approx  r_{\rm co}=2\cdot 10^8\,$cm),
 would be very luminous in hard X-rays 
but relatively little mass were accreted onto the neutron star (so that the
torque be low).
One difficulty with such a scenario is purely empirical:
 the effective temperature \citep{SS73} of such a disc,
$T\approx [L_{\rm D}/(\sigma\pi r_{\rm co}^2)]^{1/4}$ should be one third
the value of the temperature
for an Eddington luminosity disc extending to the surface
of the neutron star, i.e., no more than 1 keV. 
The large flux observed by Nu-STAR in the 3-30 keV range
makes this unlikely. 

Further, to have 
most of the luminosity coming from the disc with
an inner edge at approximately $r_{\rm co}=2\cdot 10^8\,$cm,
one would have to have
a model in which the mass accretion rate\footnote{According 
to  the \citet{SS73}  model  \citep[see also][]{King09} of super-Eddington
discs, the luminosity increases only logarithmically with the mass accretion
rate. If this model were applicable to the current
situation, $\dot M_{\rm D}$ would have to be much higher than
$10^{-4}{\rm M_\odot}$/y. On the other hand, this model neglects advection of
energy which may play a crucial role in the accretion rate radial distribution
 \citep[e.g.,][]{Sadowskietal14}.} 
through the disc terminating at 
$\sim 100$
stellar radii from the neutron star,
$\dot M_{\rm D}\gtrsim   L_{\rm D}r_{\rm co}/(GM)$,
would have to be
 $>10^4$ the Eddington value
 (for $L_{\rm D}\approx 10^2 L_{\rm Edd}$), corresponding to a mass transfer
rate from the companion $>10^{-4}{\rm M_\odot}$/y.
 
In view of these difficulties, we will start our discussion of the spin-up
torques by examining the most conservative model of NuSTAR J095551+6940.8,
 a disc extending essentially to the surface of the neutron star.
\section{spin-up of an ultraluminous weakly magnetized neutron star}
\label{s3}

Neglecting the dynamical influence of magnetic fields
(dipole magnetic field at the stellar surface of
$\sim 10^9\,$G or less), one expects the accretion disc
in a semi-detached neutron star binary
to extend to the surface of the star or to terminate between the
marginally stable and marginally bound orbit, depending on the accretion
rate and the compactness of the star. 

Let us estimate the expected isotropic emission $L_X$ associated with
the spin-up rate of the neutron star in such
a situation, assuming 
canonical neutron star parameters: mass $M=M_0\equiv1.4 \,{\rm M_\odot}$,
radius $R=R_0\equiv10\,$km, and moment of inertia 
$I=I_0\equiv 10^{45}\,{\rm  g\cdot cm^2}$.
 In the next section we will place constraints on
these parameters derived from the observed properties of M82 X-2.

For a slowly spinning pulsar, the Schwarzschild metric is an excellent
approximation, and as the marginally stable orbit (ISCO) is at
$r_{\rm ms}=6GM_0/c^2 = 13\,{\rm km}>R_0$, we can take the rate
of accretion of angular momentum to be 
\begin{equation}
\label{eq:jzero}
\dot J_0\equiv\dot M c\times r_{\rm ms}/\sqrt{3},
\end{equation}
assuming Keplerian motion in the marginally stable orbit.
Here, and elsewhere, $\dot M$
is the mass accretion rate onto the neutron star.
The related spin-up rate is
\begin{equation}
\label{eq:spinup}
\dot \nu_0 =\frac{\dot J_0}{2\pi I_0},
\end{equation}
Taking $\dot \nu_0 = \dot \nu= 10^{-10}\,{\rm s^{-2}}$,
we obtain 
$\dot M=0.3\times10^{20}\,{\rm g/s}\approx0.5\times10^{-6}\,{\rm M_\odot}/$y
corresponding to the luminosity $L_0=GM\dot M/R_0=0.5\times10^{40}\,$erg/s,
which compares very favorably with the (isotropic) luminosity inferred
from the observations $L_X=10^{40}\,$erg/s.
We have ignored the modest redshift correction
to $L_0$, of magnitude $[1-2GM_0/(R_0c^2)]\sim 0.7$,
since about one half of the power is released at larger radii in the accretion
disc. 
Thus, we see that the observed luminosity is a direct consequence
of mass accretion onto the surface of a neutron star at the rate necessary
to explain the spin-up rate with a lever arm approximately equal to the
stellar radius. 

\section{Nature of the ultraluminous pulsar M82 X-2}
In the previous section we used canonical values for the neutron star
parameters to infer that there is no evidence of a strong magnetic field
in the spin-up behaviour of the  M82 X-2 source,
and no compelling reason to expect the accretion torque to change in the
future.

Had we assumed a strong magnetic field
that terminates the disc far above the stellar surface
 (i.e., at $r\gtrsim 100 R_0$ for $B\gtrsim10^{12}\,$G), 
the mass accretion rate corresponding to the same torque
would have been lower
by a factor $\sim\sqrt{r_{\rm ms}/r}< 0.1$ decreasing the luminosity
released at the surface by a factor of 10. It is hard to see how
this deficit in hard X-ray luminosity could be made up by the disc.
As remarked in Section~\ref{s2}, at such large distances from
the neutron star, the disc should be emitting soft X-rays, at most.
In other words, if   the disc terminates far above the stellar surface, 
the mass accretion rate inferred from the luminosity
would have been the same
(assuming isotropic emission), but the torques would
be enhanced by a factor $\sim\sqrt{r/r_{\rm ms}}> 10$, leading to a spin-up
rate much larger than the observed one.
One could argue that the accretion torques could be compensated
by magnetic torques transmitting angular momentum back to the
accretion disc, but it would be an unexplained coincidence that
the difference of two larger torques results in a value exactly matching
the one corresponding to $r\approx R$.

So far we have ignored possible beaming of radiation.
 We now allow this possibility with no theoretical
prejudice as to its origin.
We would like to examine the constraints on
the compact source imposed by the observations of the M82 X-2
pulsar. The period alone,  $P=1.37\,$s rules out white dwarfs and
less compact stars \citep{ThorneIpser68}, and imposes a lower limit to the
mean density of the compact object of $\bar\rho>10^8{\rm g/cm^{3}}$,
following from the mass-shedding limit $\sqrt{4\pi G\bar\rho/3}>2\pi/P$.
The coherent periodicity obviously rules out black holes.
Of the astronomical objects known so far, only neutron stars are compatible
with the measured period. 

The question is whether we can determine the mass and radius
of the (presumed) neutron star, as well as the accretion geometry.
The simultaneous measurement of the luminosity of the source and
its spin-up is quite constraining, although the radius cannot
be determined separately from the mass at present.

For convenience we will eventually parametrize the effective 
radius $r$ corresponding to the torque lever arm with
the radius of the marginally stable orbit in the Schwarzschild
metric, $r_{\rm ms}=6GM/c^2$. Models of neutron stars give stellar radii very
close to that value for most equations of state of dense matter
\citep{KW85,Cooketal94}, so $r/r_{\rm ms}\approx r/R$.
The luminosity expression will differ from the isotropic one by
a beaming factor, $b\le 1$ and we will allow redshift corrections
$f_1(R)\sim 1$,
\begin{equation}
\label{eq:lx}
b L_X=f_1(R)GM\dot M/R.
\end{equation}
Here, $b L_X$ is the true luminosity. 
In addition to the ``material'' torque corresponding to the advected
angular momentum proportional to the product of $\dot M$ and
of the specific angular momentum of the accreting matter, magnetic
torques may be present. The latter may be of either sign, depending
on the accretion geometry which is related to $\dot M$ and the
 magnetic dipole moment. Indeed, observations of X-ray pulsars
near equilibrium
reveal periods of spin-up followed by periods of spin-down at comparable
rates \citep{Bildstenetal97}. We are going to parametrize the total torque
acting on the neutron star by an effective radius corresponding to the total
torque divided by the orbital specific angular momentum at that radius
times $\dot M$. 
\begin{equation}
\label{eq:angmrate}
\tau= f_2(r)\sqrt{GMr}\dot M,
\end{equation}
with $f_2(r)$ describing relativistic corrections.
The specific angular momentum value, $J_0$,
for test particles in the marginally stable orbit is recovered with 
$r=r_{\rm ms}$ and $f_2(r)=\sqrt{2}$.
The moment of inertia of the neutron star can be written as
\begin{equation}
\label{eq:mominert}
I=\beta MR^2,
\end{equation}
with $\beta \approx 0.3$ \citep{Urbanecetal13}.
Applying the torque to the star, 
\begin{equation}
\tau=2\pi\dot\nu I,
\end{equation}
we obtain
\begin{equation}
MR=
\frac{f_2(r)}{\beta c f_1(R)}\frac{bL_X}{2\pi\dot\nu}\sqrt{\frac{rc^2}{GM}}.
\end{equation}
Putting $f_2/(\sqrt6\beta f_1)\approx 1$
one gets
\begin{equation}
\label{eq:btwo}
\frac{ b^2 r}{ r_{\rm ms}}\approx 
 \left(\frac{R}{R_0}\right)^2 
\left(\frac{\dot\nu\times10^{50}\rm erg\cdot s}{ L_X}\right)^{2}
\left(\frac{M}{1.4\,{\rm M_\odot}}\right)^2.
\end{equation}
Several conclusions can be drawn from this equation, which can also
be written in the form
$$\frac{rr_{\rm ms}}{R^2}\approx
  \left(\frac{\dot\nu \times 10^{50}\rm erg\cdot s}{bL_x}\right)^{2}
\left(\frac{M}{1.4\,{\rm M_\odot}}\right)^4.$$

First, if there is no strong beaming, i.e., $b\sim 1$, applying
an accretion torque corresponding to an inner disc radius far above
the stellar surface, $r\gg R$, would imply a large mass of the compact
star, requiring an exotic model of dense matter.
In the standard magnetized accretion powered pulsar models,
the value of $r$ is very close to the co-rotation radius.
A constraint on $r$ can then be translated into a co-rotation
frequency, i.e., equilibrium rotation frequency of the pulsar.
The largest measured (and already challenging nuclear theorists)
neutron star mass \citep{Demorestetal10,Antoniadis13}
 is $M=2 \,{\rm M_\odot}$, for which all theoretical models give $R<r_{\rm ms}$.
Hence, for standard
neutron stars we have an upper limit of
$r<5r_{\rm ms}$, corresponding to a corotation frequency of
$ >100\,{\rm Hz}\times(2 \,{\rm M_\odot}/M)$.

Second, a small value of the beaming factor, $b\ll 1$ could in principle
allow large values for the inner radius of the accretion disc, 
$r\sim b^{-2}r_{\rm ms}\approx  b^{-2}R$,
but the beaming could not be then provided by the accretion funnel postulated
in other ULXs (e.g., King et al. 2001), 
the disc being here too far from the compact star to
collimate the radiation. 
If one wanted to invoke
beaming to reduce the luminosity of the source inferred from the observed
flux, it would have to occur near the surface of the star, and yet
allow for a large part of the flux to be unmodulated by the rotation
of the neutron star.

One could invoke collimation by the magnetic
field tube at the pole \citep[perhaps a fan beam
caused by radiation escaping sideways from the accretion column
in a strong magnetic field,][]{Gnedin}. A beaming fraction $b\sim 10^{-1}$ 
would increase the magnetic truncation radius of the disc
by the requisite factor of 100, and decrease the mass accretion rate 
inferred from the luminosity by a factor of 10.
For the system at hand, this would imply a mass accretion rate
$b\times10^{-6}{\rm M}_\odot/{\rm y}=10^{-7}{\rm M}_\odot/{\rm y}$
\citep[which could be hard to understand in a system presumably
undergoing unstable mass transfer through the Roche lobe, e.g.,][]{Tauris11}.
However, to be consistent one should do the same
for the ordinary X-ray pulsars, as it would otherwise be hard to see why
increasing the luminosity should lead to increasingly   beamed
emission (lower values of $b$). For the ordinary X-ray pulsars
the same beaming fraction, $b=0.1$, would necessitate increasing the magnetic
radius to the  impossibly
large value $\approx 10^{10}{\rm cm}$,
and decrease the typical mass transfer rate to $10^{-11}\,{\rm M_\odot}/{\rm y}$.

Further, there remains an improbable coincidence
related to the one mentioned at the end of the second paragraph of this section. 
We are unaware of any mechanism 
which would relate the beaming factor to the inverse square root of
the magnetic or corotation radius, as required by Eq.~(\ref{eq:btwo}):
$b\approx(r_{\rm ms}/r)^{1/2}$.

\subsection{An ultraluminous accreting millisecond pulsar}
In view of the difficulties associated with the strongly magnetized
pulsar model for NuSTAR J095551+6940.8, 
and the coincidences required to make it compatible with the observations,
it seems much more natural to assume that $b\sim1$, implying that
$M\approx1.4\,{\rm M_\odot}$, and that the inner disc radius is
comparable to the stellar radius $r\sim R\approx r_{\rm ms}$, with no
strong upper limits on the future spin frequency of the M82 X-2
neutron star.

The idea that accreting weakly magnetized neutron stars may exhibit
pulsations in X-rays is not new, with the first such system, the 400
Hz SAX J1808.4-3658 pulsar, discovered more than a decade ago
\citep{WvdK98}.  Of course, the known accreting millisecond pulsars
are not HMXBs, in fact the typical companion mass is
$\lesssim0.1\,{\rm M_\odot}$. We take it as an empirical fact that accreting
   weakly magnetized neutron stars become X-ray pulsars under some
   conditions, presumably channelling a large part of the accretion
   flow to the magnetic poles.  What these conditions need to be is an
   open  theoretical question.  Many LMXBs have accretion rates which
   are a large fraction of the Eddington limit value, and some of them
   exhibit transient pulsations
   \citep{Galloway07,Altamirano08,Casella08}. At present, it is not
   known why millisecond accreting pulsars do not always exhibit their
   periodicity (pulsations), and whether this is related to the mass
   accretion rate. In particular, it is not known whether rapid
   rotation plays a role in the suppression of pulsations, i.e.,
   whether or not weakly magnetized neutron stars are more likely to
   exhibit pulsations when rotating slowly.

Assuming that our suggestion of low magnetic field in the source is correct,
is there any way for NuSTAR J095551+6940.8 to avoid being spun up
to millisecond periods? The neutron star could not yet have
accreted more than a small percentage of a solar mass
(about $10^{-3}{\rm M_\odot}$ judging by its current period of 1.37 s),
and will not accrete more than $0.1{\rm M_\odot}$ before it is spun up
to hectoHz frequencies. So unless it was formed as an unusually massive
($2{\rm M_\odot}$) object, the neutron star is not going to collapse to
a black hole before it becomes a millisecond pulsar.

One difficulty with our model of persistent spin-up is that we
   seem to be observing the pulsar in the first $10^3\,$y of a
   $10^5\,$y accretion episode, which on the face of it would
   correspond to one chance in a hundred. In view of the theoretical
   uncertainties discussed in the previous two paragraphs it is hard
   to reliably estimate the selection effects involved, i.e., over
   what fraction of the $10^5\,$y epoch of mass transfer the pulsar
   would be detectable (or even exist at all). 

Another possible way to avoid spin up is to turn off accretion onto
the star. This seems unlikely. With a companion so massive,
and such a high luminosity at this orbital period
the system is probably undergoing unstable mass transfer through
Roche-lobe overflow which is going to continue until  the companion
loses most of its mass. Of course, since the discovery of the first
eclipsing pulsars we know of systems where mass transfer (loss from the
companion) occurs with no accretion onto the neutron star,
however this is an unlikely scenario for the binary at hand.
Although, perhaps, at some point in time the pulsar could turn on in the radio,
(i.e., become a radio pulsar if the mass transfer rate temporarily drops)
as predicted theoretically \citep{eclipsing} and recently observed
\citep[e.g.,][]{2009Sci...324.1411A,bassa},  it would take a very powerful radio
pulsar indeed to expel from the system
matter transferred at such a prodigious rate. Note that
the known eclipsing pulsars that ablate their companions are in fact millisecond
pulsars, and their companions have fairly low masses.

\subsection{Alternative pulsar models}
Several authors suggest a different model.
  \cite{Bachettietal14,Christo14,Ekcietal14}; and \cite{Lyutikov14}
  assume that the M82 X-2 pulsar is close to spin equilibrium, and
  obtain $\sim1-100\,$TG values for the magnetic field $B$. While this
  cannot be excluded with the present data (notwithstanding the
  improbable coincidences discussed above), we note that the
  assumption of equilibrium spin is notoriously unreliable as a
  predictor of magnetic dipole strength in massive binaries. 

Several Be X-ray binaries exhibit observed cyclotron lines
   corresponding to a 3 TG field, and yet their periods span 3.5
   orders of magnitude \citep{Klus14}.  If assumed to be in torque
   equilibrium they should all have the same period, as the adopted
   torque model gives a one to one correspondence between the magnetic
   field strength and the equilibrium spin (the luminosity does not
   vary much from source to source in the sample). In other words,
   estimates of the magnetic field in the same sample, if based on the
   assumption that the observed period is the equilibrium one, would
   lead to the predicted $B$-field value being up to 3 orders
   of magnitude larger than the actually measured values  \citep{Klus14}.

\section{Past and future evolution of the ultraluminous pulsar}

The mass of the companion star of the M82 X-2 pulsar is (assuming the
neutron star mass
 $M=1.4 \,{\rm M}_{\odot}$) $M_2 >5.2\, {\rm  M}_{\odot}$
 \citep{Bachettietal14}. To provide the required mass
accretion rate $\dot M \ga 10^{-6} {\rm M}_{\odot}$/y the companion
has to fill its Roche lobe. Since the ratio of the companion to the
neutron star masses is $q\equiv M_2/M >1$ the mass from the donor
will be transferred on its thermal timescale assuming its envelope is
radiative. The evolution of such systems has been studied by
\citet{1999MNRAS.309..253K} \citep[see also][]{Kingetal00}.

If the mass of the companion is not much higher than the lower
  limit of  $5.2\, {\rm  M}_{\odot}$, the past evolution of the system
  could be one considered for intermediate mass X-ray binaries (IMXBs).
  Generally speaking, no detailed calculation of the magnetic field
  origin in supernova explosions are available. Perhaps the different
  evolutionary scenario of HMXBs and of ULXs could be responsible for
  a lower field value in the M82 X-2 case than in the common HMXBs.
  Until now IMXBs were a theoretical concept thought unlikely to be
  observed because of the fairly brief fraction of their lifetime
  corresponding to episodes of high mass transfer. However,
  observation of a ULX selects this very phase of rapid mass transfer.
  In particular, Case A of Roche lobe overflow of \cite{Tauris11}
  corresponds to a $\sim10^5\,$y epoch of mass transfer in excess of
  $10^{-6} {\rm M}_{\odot}$/y. The progenitor would have been a $\ge20
  {\rm  M}_{\odot}$ primary with a $\ge 5{\rm  M}_{\odot}$ secondary,
  the high mass transfer phase occcuring $\sim100$ million years after
  formation, following a common envelope phase and a supernova
  explosion, with the system evolving towards a neutron star - white
  dwarf binary \citep{Tauris11}, a perfect progenitor for a ``black
  widow'' pulsar ablating its companion
  \citep{Fruchter88,eclipsing,Eichler88,KCR92}, if the pulsar is ever
  spun-up to a millisecond period.

 However, because
of the relatively high donor mass $M_2>5\, {\rm M}_{\odot}$ the future
of the M82 X-2 binary is difficult to predict since it is unlikely
that it will avoid the the so-called ``delayed dynamical instability"
\citep[e.g.,][]{Webbink77} when the companion star expands
adiabatically in response to mass loss. In principle this should force
the system to go through a second common envelope phase whose outcome is
notoriously uncertain \citep[][]{Ivanovaetal13}. On the other hand
\citet{1999MNRAS.309..253K} notice that ``even very rapid mass
transfer on to a neutron star does not necessarily result in a common
envelope."

Therefore the future fate of the M82 X-2 accreting pulsar is also uncertain, and
depends on the the mass loss and angular momentum loss from the system.
Mass transfer of only $0.2 M_\odot$ is in principle
sufficient to spin up a weakly magnetized pulsar to
800 Hz, i.e., faster than any observed one, and of $0.3 M_\odot$
to the mass-shedding limit. If mass transfer continues, the pulsar must
lose angular momentum. If it is unable to transfer
angular momentum outwards through the disc, or to lose it to a jet,
it may become a powerful source of gravitational
radiation \citep{Wagoner84},
 until such time as mass accretion stops or, alternatively,
the neutron star will collapse to a black hole. 
In the latter case a period of $\sim10^5\,$y of powerful steady emission
of gravitational waves would be followed by black hole
formation outside a supernova, in an accretion induced collapse,
which could lead to a black hole ULX or, depending on the mass ratio
of the components at the time of the collapse,
possibly unbind the binary yielding a fairly low
mass black hole which could escape the galactic
plane, as would the erstwhile companion.

\section{Conclusions}

 A neutron star in an intermediate (IMXB) or
a high mass X-ray binary (HMXB) will be spun up
to millisecond periods upon accreting about $0.1\,{\rm M_\odot}$, if only its magnetic
dipole moment is sufficiently low to allow the accretion disc 
to extend to the stellar radius, or its immediate vicinity.
Judging by its apparent luminosity and the measured
value of its spin-up, this seems to be the case for the M82
X-2 source, the  $1.37\,$s pulsar NuSTAR J095551+6940.8.

In subsequent phases
of its evolution the pulsar must either avoid further accretion 
of angular momentum or become
a powerful source of gravitational radiation. The neutron star may either
survive as a millisecond pulsar or collapse to a black hole depending on
how effective it is in expelling most of the matter transferred from the massive
companion. 

Whether or not our suggestion turns out to be true, one needs to entertain
the possibility that other ULXs may be pulsars. 
For instance, the recently reported
3.3 Hz and 5 Hz frequencies in
the X-ray flux of M82 X-1 \citep{Pasham14}
could in fact turn out to be harmonics of $1.67$ Hz, i.e., that source may
be a $P=0.6$ s pulsar. 
Another ULX, in NGC 7793, has an upper limit to the mass
of the compact object of $<15M_\odot$, with allowed solutions in the neutron star
mass range and phenomenology of state transitions similar to that of Her X-1
\citep{Motch14}.

\section*{Acknowledgments}
We thank the anonymous referee for the very helpful comments, and
Thomas Tauris for discussion.
This work was supported in part by the Polish NCN grant UMO-2013/08/A/ST9/00795.
JPL acknowledges support from the French Space Agency CNES.

\label{lastpage}

\end{document}